%% file: gengsf.tex
\documentclass[
	aps,
	prd, 
	twocolumn, 
	a4paper, 
	10pt, 
	amsmath,
	amssymb, 
	preprintnumbers,  
	tightenlines, 
	notitlepage,
	floatfix,
	nofootinbib,
	longbibliography
	]{revtex4-1}\pdfoutput=1\pdfsuppresswarningpagegroup=1
\usepackage[utf8]{inputenc}
\usepackage[english]{babel}
\usepackage{bbm}
\usepackage{graphicx}
\usepackage{hyperref}
\usepackage{wasysym}
\usepackage{amsthm}
\usepackage{dcolumn}
	\newcolumntype{.}{D{.}{.}{13}}
	\newcolumntype{d}[1]{D{.}{.}{#1}}
\usepackage{color}	
\usepackage[all,cmtip]{xy}
\usepackage{tensor}

\graphicspath{{pics/}}

\allowdisplaybreaks[0]
\input{commands.input}
\begin{document}

\title{Gravitational self-force on generic bound geodesics in Kerr spacetime}

\author{Maarten \surname{van de Meent}}
\email{mmeent@aei.mpg.de}
\affiliation{Max Planck Institute for Gravitational Physics (Albert Einstein Institute), Potsdam-Golm, Germany}
\affiliation{Mathematical Sciences, University of Southampton, United Kingdom}

\date{\today}
\begin{abstract}
In this work we present the first calculation of the gravitational self-force on generic bound geodesics in Kerr spacetime to first order in the mass-ratio. That is, the local correction to equations of motion for a compact object orbiting a larger rotating black hole due to its own impact on the gravitational field. This includes both dissipative and conservative effects. Our method builds on and extends earlier methods for calculating the gravitational self-force on equatorial orbits. In particular we reconstruct the local metric perturbation in the outgoing radiation gauge from the Weyl scalar $\psi_4$, which in turn is obtained by solving the Teukolsky equation using semi-analytical frequency domain methods. The gravitational self-force is subsequently obtained using (spherical) $l$-mode regularization.

We test our implementation by comparing the large $l$-behaviour against the analytically known regularization parameters. In addition we validate our results be comparing the long-term average changes to the energy, angular momentum, and Carter constant to changes to these constants of motion inferred from the gravitational wave flux to infinity and down the horizon.

\end{abstract} 
%\pacs{4.25.Nx,4.30.Db,45.50.Pk}

\maketitle
\setlength{\parindent}{0pt} 
\setlength{\parskip}{6pt}

\section{Introduction}
For the interpretation of gravitational wave observations, accurate theoretical models of their sources are essential. For the comparable mass binaries observed by LIGO and Virgo \cite{GW150914,GW151226,GW170104,GW170814,GW170817} this modelling is provided  by the results of post-Newtonian (PN) theory and numerical relativity (NR), typically repackaged in an effective-one-body~(EOB) model  or a phenomenological surrogate model. However, PN theory works well only in the (relative) weak field regime, whereas NR simulations are limited to systems with fairly homogeneous intrinsic length scales, practically limiting its applicability to systems with small mass-ratios ($\mr := m_2/m_1\gtrsim 1/10$). Consequently, current methods are insufficient to accurately model the final strong field stages of the inspirals of small mass-ratio binaries.

Nonetheless, the current generation of ground based detectors is in principle sensitive to binaries with mass ratios as low as $10^{-2}$, which currently cannot be accurately modelled. The occurrence of such small mass-ratio inspirals is dependent on the existence of a sufficiently large population of large $\sim 100 M_{\astrosun}$ black holes, which is not guaranteed to exist. Hence LIGO/Virgo observation of such small ratio events is not a given.

ESA's planned space-based gravitational wave observatory, LISA, will however be sensitive to so called extreme mass ratio binaries (EMRIs), compact binaries consisting of a $\sim 10^6 M_{\astrosun}$ supermassive black hole and a stellar mass compact object. The rate at which EMRIs occur is uncertain, but studies show we should expect between 1 and 4000 detectable LISA events per year with SNRs up to a few hundred \cite{Babak:2017tow}.

Unlike the comparable mass binaries detectable by ground based detectors, EMRIs are expected to exhibit significant eccentricity ($e \lesssim 0.2$ at merger), and inclination of the orbital plane compared to the total angular momentum. Moreover, the small mass ratio implies that evolution of these systems is very slow, producing $\sim \mr^{-1} \simeq 10^5$ gravitational wave cycles in the strong field regime. As a consequence, EMRIs produce an information rich GW signal, allowing highly accurate determination of the source properties. The component masses, primary spin, eccentricity and inclination can be determined at a relative accuracy of $10^{-5}$, while luminosity distance can be determined to $5-10\%$, and sky position can be localized to a few square degrees \cite{Babak:2017tow}. Alternatively, the detailed signal can be used to test the general relativity prediction that the supermassive primary should be described by the Kerr metric, by measuring its mass quadrupole to a relative accuracy of $10^{-4}$ \cite{Babak:2017tow}.

However, any such measurement will rely on the availability of accurate waveform models for EMRIs including the effects of spin, eccentricity, and inclination. One approach is to use the smallness of the mass-ratio $\mr$ to our advantage, and treat dynamics of EMRIs in a systematic perturbative expansion in $\mr$. At zeroth order in $\mr$, the secondary object acts as a test particle in the Kerr geometry generated by the primary. It follows a geodesic, which can be obtained analytically \cite{Fujita:2009bp,Hackmann:2008zz,Hackmann:2010tqa,Hackmann:2010zz}. At the next order, the corrections to the equations of motion due to the gravitational field generated by the secondary can be collected into an effective force term perturbing the geodesic equation, the gravitational self-force (GSF). Since this force is small the evolutionary timescale ($t_\mathrm{insp} = \bigO(\eta^{-1})$) of an EMRI is much larger than the orbital timescale ($t_\mathrm{orb} = \bigO(1)$). This hierarchy of timescales can be exploited to simplify the evolution of EMRIs by using a two timescale expansion. A systematic analysis by Hinderer and Flanagan \cite{Hinderer:2008dm} has shown that in order to obtain the phase evolution of an EMRI with error of $\bigO(\eta)$ we need the first order GSF sourced by individual geodesics, and the long term average of the dissipative part of the second order GSF. In this paper we will provide the first calculation of the first order GSF on fully generic bound geodesics featuring both eccentricity and inclination in Kerr spacetime.

The formalism for calculating the GSF was first introduced by Mino, Sasaki, and Tanaka \cite{Mino:1996nk} and Quinn and Wald \cite{Quinn:1996am} in the mid 1990s. In the two decades since, the formalism has been further refined improving both mathematical rigour and conceptual clarity (see \cite{Poisson:2011nh,Pound:2015tma} for reviews and references).

Numerical calculations of the GSF have made steady progress over these last two decades. The first numerical calculations appeared in 2002 for direct radial plunges into a Schwarzschild black hole \cite{Barack:2002ku}. The calculation of the GSF on circular orbits followed in 2007 \cite{Barack:2007tm}, and completely generic bound eccentric orbits in 2009 \cite{Barack:2009ey}. By now, first order GSF calculations in Schwarzschild spacetimes are routine, using a wide variety of numerical methods, regularization techniques, and gauges \cite{Sago:2008id,Detweiler:2008ft,Shah:2010bi,Akcay:2010dx,Barack:2010tm,Hopper:2010uv,Dolan:2012jg,Akcay:2013wfa,Osburn:2014hoa,Merlin:2014qda,Wardell:2015ada,vandeMeent:2015lxa,vandeMeent:2016pee}. The results of these calculations have been used to evolve EMRIs around a Schwarzschild black hole \cite{Warburton:2011fk,Osburn:2015duj}

For a long time, calculating the GSF on a Kerr background remained a challenge. The core obstacle was that the linearized Einstein equation on a Kerr background cannot be separated by introducing some set of harmonics. This has led Dolan and Barack to pursue 2+1 dimensional techniques for calculating the GSF \cite{Dolan:2010mt,Dolan:2011dx,Dolan:2012jg}. However, these methods suffer from numerical instabilities that have been overcome to produce the GSF on circular equatorial orbits \cite{Dolanunpublished}, but thus far have prevented their application to more general orbits.

Another approach that had been considered, is to utilize the fact that the Weyl scalars $\psi_0$ and $\psi_4$ satisfy the (separable) Teukolsky equation \cite{Teukolsky:1972my,Teukolsky:1973ha}, while containing most of the gauge invariant information about the full metric perturbation \cite{Wald:1973}. In the 1970s, Chrzanowski, Cohen, and Kegeles \cite{Cohen:1974cm,Chrzanowski:1975wv,Kegeles:1979an} developed a method for reconstructing vacuum metric perturbations in a radiation gauge from vacuum solutions of $\psi_0$ or $\psi_4$. However, as noted by Ori \cite{Ori:2002uv}, when this procedure is applied to a field sourced by a point particle the resulting metric perturbation is highly singular. Not only does the resulting metric perturbation feature a singularity at the location of the particle, but in addition a string-like gauge singularity extends from the particle to black hole horizon and/or infinity. It was unclear whether the established GSF formalism would extend to such singular gauges. Only in 2013 did Pound, Merlin, and Barack \cite{Pound:2013faa} show that the GSF can be extracted from radiation gauge metric perturbations.

A second issue was that while $\psi_0$ or $\psi_4$ contain most information about the metric perturbation, they are oblivious to perturbations within the Kerr family of metric solutions \cite{Wald:1973}. These ``mass'' and ``angular momentum'' perturbations need to be recovered through other means. Merlin et al.\cite{Merlin:2016boc} recovered these pieces for fields sourced by a particle on an equatorial orbit by imposing continuity of certain gauge invariant fields constructed from the metric. The result is remarkably simple; in the region ``outside'' the particle orbit the mass and angular momentum perturbations are given simply by the energy and orbital angular momentum of the orbit, while both perturbations vanish ``inside'' the orbit. By directly analyzing the form of metric perturbations resulting from the CCK procedure, it was shown in \cite{vandeMeent:2017fqk} that this result must in fact hold for any source with compact support in the radial direction.

Pending resolution of both issues, implementation of the radiation gauge approach to calculating the GSF was pioneered by the group of Friedman \cite{Keidl:2006wk,Keidl:2010pm,Shah:2010bi}, culminating in the calculation of the Detweiler redshift invariant on circular equatorial orbits in Kerr spacetime \cite{Shah:2012gu}. In previous papers \cite{vandeMeent:2015lxa,vandeMeent:2016hel}, the author expanded on their techniques to obtain the first order GSF and redshift on eccentric equatorial orbits. In this paper, we will tackle the case of generic bound orbits in Kerr spacetime, featuring both eccentricity and inclination.

The plan of this paper is as follows. In Sec. \ref{sec:prelim} we review the preliminaries necessary for our calculation. Section \ref{sec:method} then reviews our method for calculation of the GSF in radiation gauge using a metric reconstructed from $\psi_4$. In this section, the focus will be on aspects of the method that change for generic Kerr geodesics. A selection of results is presented in Sec. \ref{sec:results}, going through various consistency checks and plots of the final results. We conclude with a discussion of ways these calculations can be used to explore new physics.

\subsection{Conventions}
This paper uses an overall metric signature of $(-+++)$; for further sign conventions regarding the definitions of other quantities such as the Weyl curvature scalars we use conventions consistent with Appendix A of \cite{vandeMeent:2016hel}. We further work in geometrized units such that $(c=G=M=1)$.

\section{Review of preliminaries}\label{sec:prelim}
In this section we review some of the preliminaries needed for a calculations. Along the way we establish some of the notations and conventions used.
\subsection{Generic geodesics in Kerr spacetime}
Like in previous papers \cite{vandeMeent:2015lxa,vandeMeent:2016hel}, we consider the Kerr metric in modified Boyer-Lindquist coordinates, where the polar angle $\theta$ has been replaced by $z:=\cos\theta$. In these coordinates the Kerr metric generated by a black hole with mass $M=1$ and spin $a$ is given by
\begin{equation}
\begin{split}
\label{eq:kerr}
\id{s}^2 = 
-\bh{1 - \frac{2r}{\Sigma}}\id{t}^2 
+ \frac{\Sigma}{\Delta} \id{r}^2
+ \frac{\Sigma}{1-z^2} \id{z}^2
\\
+ \frac{1-z^2}{\Sigma} \bh{2a^2 r (1-z^2)+(a^2+r^2)\Sigma}\id\phi^2
\\
- \frac{4ar(1-z^2)}{\Sigma}\id{t}\id\phi,
\end{split}
\end{equation}
with
\begin{align}
\Delta &= r(r-2) + a^2,\\
\Sigma &= r^2 + a^2 z^2.
\end{align}

At zeroth order an object with mass $m<<M=1$ follows a geodesic in the Kerr background,
\begin{equation}
m\d{p^\mu}{\pt}+ \Gamma^{\mu}_{\alpha\beta}p^\alpha p^\beta =0,
\end{equation}
where $p^\mu := m u^\mu = m\d{x^\mu}{\pt}$ is the four-momentum, $\pt$ is proper time, and $\Gamma^{\mu}_{\alpha\beta}$ are the Christoffel symbols of the Kerr metric.

Solving the geodesic equation in Kerr spacetime is greatly helped by the existence of a complete set of constants of motion. The first is the invariant mass $-m^2 =p^\mu p_\mu$. Furthermore, the Kerr metric \eqref{eq:kerr} has two explicit symmetries expressed by the Killing vectors $(\pd{}{t})^\mu$ and $(\pd{}{\phi})^\mu$, which give rise to two further constants of motion; the specific energy $\nE:=-u_\mu (\pd{}{t})^\mu$, and the specific (orbital) angular momentum $\nL:=u_\mu (\pd{}{\phi})^\mu$. Finally, Carter showed \cite{Carter:1968rr} that the Kerr metric has a third hidden symmetry expressed by a Killing tensor,
\begin{equation}
K_{\mu\nu} :=2 \Sigma l_{(\mu}n_{\nu)} +r^2 g_{\mu\nu},
\end{equation}
where are $l^\mu$ and $n^\nu$ are the principal null vectors of the Kerr metric,
\begin{align}
l^\mu &:= (\frac{r^2+a^2}{\Delta},1,0,\frac{a}{\Delta})\text{, and}\\
n^\mu &:= (\frac{r^2+a^2}{2\Sigma},-\frac{\Delta}{2\Sigma},0,\frac{a}{2\Sigma}).
\end{align}
This Killing tensor defines a fourth constant of motion, the Carter constant,
\begin{equation}
\nQ := u^\mu K_{\mu\nu} u^\nu - (\nL-a\nE)^2
\end{equation}

Using this complete set of constants of motion, the equations of motion for a geodesic can be rewritten,
\begin{subequations}\label{eq:eom0}
\begin{align}
\Bh{\Sigma\d{r}{\pt}}^2 &= \hh{\nE(r^2+a^2)-a\nL}^2\\
&\qquad-\Delta\hh{r^2 +(\nL-a\nE)^2 +\nQ},\notag \\
\Bh{\Sigma\d{z}{\pt}}^2 &= a^2 (1-\nE^2)z^4\\
&\qquad-\hh{\nQ+a^2(1-\nE^2)+\nL^2}z^2+\nQ,  \notag\\
\Sigma\d{\phi}{\pt} &=\frac{a}{\Delta}\hh{\nE(r^2+a^2)-a\nL}+\frac{\nL}{1-z^2},  \\
\Sigma\d{t}{\pt} &=\frac{r^2+a^2}{\Delta}\hh{\nE(r^2+a^2)-a\nL}-a^2\nE (1-z^2).  
\end{align}
\end{subequations}
By introducing the Mino time parameter $\mt$ defined by
\begin{equation}
 \d{\pt}{\mt}=\Sigma,
\end{equation}
the radial and polar motions can be completely decoupled. Geodesic motion around a Kerr black hole can therefore be viewed as two completely independent motions (radial and polar). This inspires us to introduce two periodic phase coordinates $q_r$ and $q_z$, which specify where along each of the cycles the particle is. We further specify that both phases evolve linearly with Mino time
\begin{align}
 \d{q_r}{\mt} &= \U_r\text{, and} \\
 \d{q_z}{\mt} &= \U_z, 
\end{align}
where $\U_r$ and $\U_z$ are the frequencies with respect to Mino time of the radial and polar motions, explicit expressions for which can be found in \cite{Fujita:2009bp}. We further adopt the convention that $q_r = 0$ corresponds to the apapsis of the radial motion, meaning that $q_r = \pi$ will correspond to periapsis. Similarly, we choose $q_z=0$ to coincide with the polar motion reaching its maximum, which results in the minimum being reached at  $q_z=\pi$ while the equator $z=0$ is crossed at $q_z=\pi/2$ and $\q_z=3\pi/2$.

With these phases, the solutions to the Eqs. \eqref{eq:eom0} take the following form; $r$ is a periodic function of just $q_r$, $z$ is a periodic function of just $q_z$, $t$ and $\phi$ become
\begin{align}\label{eq:t}
t &=  \U_t\mt +t_r(q_r) +t_z(q_z)\text{, and}\\
\phi &=  \U_\phi\mt +\phi_r(q_r) +\phi_z(q_z)\label{eq:phi},
\end{align}
where $t_r$ and $\phi_r$ are purely oscillatory functions of $q_r$ and $t_z$ and $\phi_z$ are purely oscillatory functions of $q_z$.

Together with the spin $a$ the set of constants of motion $(\nE,\nL,\nQ)$ uniquely identifies a bound Kerr geodesic. However, these tend to be hard to work with. In practice, it is easier to work with a more geometric set of parameters. One such set is given by the turning points of the radial motion $\rmin$ and $\rmax$ and the turn point of the polar motion $\pm\zmax$. Here instead of  $\rmin$ and $\rmax$ we use the semilatus rectum $p$ and eccentricity $e$ defined by
\begin{align}
\rmax &= \frac{p}{1-e}\text{, and} \\
\rmin &= \frac{p}{1+e}.
\end{align}
Kerr geodesics are thus identified by a 4-tuple $(a,p,e,z)$.

\subsection{Gravitational self-force}
The main idea behind the self-force formalism is to systematically expand the equations of motion for a compact binary in powers of the small mass-ratio $\mr = m/M$. At linear order in $\mr$ to set-up is to split the metric generated by the binary as
\begin{equation}\label{eq:metricsplit}
g_{\mu\nu} +\eta h_{\mu\nu},
\end{equation}
where $g_{\mu\nu}$ is the background Kerr metric generated by the primary object, and $h_{\mu\nu}$ is a correction due to the presence of the secondary object. The motion of the secondary object is to be described by some worldline $x_0^\mu(\pt)$ in the background spacetime. This worldline is expected to satisfy a forced geodesic equation,
\begin{equation}\label{eq:gdGSF}
 m\hh{\dd{x_0^\mu}{\pt}+ \Gamma^{\mu}_{\alpha\beta}\d{x_0^\alpha}{\pt}\d{x_0^\beta}{\pt}} =\mr^2 F^\mu[h] ,
\end{equation}
where  $F^\mu[h]$ is called the \emph{gravitational self-force} or GSF.

The most rigorous approach to obtaining the key ingredients ($x_0$, $h$, and $F$) is a multi-scale expansion (see \cite{Poisson:2011nh,Pound:2015tma} for reviews). The general idea is to split the spacetime into a `near zone' where $h$ dominates the metric, and a `far zone' dominated by $g$. In each zone, the effects of the other component can be treated perturbatively. A global solution is then obtained by matching both expansions in the region where both zones overlap and both perturbative expansions hold. 

The upshot of the analysis is as follows. The world-line $x_0^\mu$ is determined by the centre-of-mass motion of the secondary (as measured asymptotically in the near-zone). The metric perturbation $h_{\mu\nu}$ is obtained by solving the linearized Einstein equation on the background $g_{\mu\nu}$ sourced by a point particle with mass $m$ following  $x_0^\mu$ and retarded boundary conditions. Finally, (if we ignore any effects from the spin of the secondary) the GSF is given by the MiSaTaQiWa \cite{Mino:1996nk,Quinn:1996am} equation,
\begin{equation}\label{eq:GSFdef}
F^\mu(\tau) =  P^{\mu\alpha\beta\gamma}\CD{\alpha}h^\reg_{\beta\gamma}(x_0(\pt)),
\end{equation}
with
\begin{equation}\label{eq:GSFprojector}
P^{\mu\alpha\beta\gamma} \equiv 
	\frac{1}{2}\hh{
		g^{\mu\alpha}u^\beta u^\gamma
		-2g^{\mu\beta}{u}^\alpha{u}^\gamma
		-{u}^\mu{u}^\alpha{u}^\beta{u}^\gamma
	},
\end{equation}
and  $h^\reg_{\mu\nu}$ is a regular part of the metric perturbation $h_{\mu\nu}$ obtained by subtracting off the Detweiler-Whiting singular field \cite{Detweiler:2002mi}. 

\subsubsection{Gauge dependence}\label{sec:gauge}
The split of the metric in Eq. \eqref{eq:metricsplit} is not unambiguous. A small change of the coordinates $x^\mu \to \tilde{x}^\mu= x^\mu +\mr \xi^\mu$ leads to a new background metric
\begin{equation}
\tilde{g}_{\mu\nu} = g_{\mu\nu} + \mr \CD{(\mu}\xi_{\nu)}.
\end{equation}
The small change can be interpreted as a part of $h_{\mu\nu}$, leading to a gauge freedom in its definition. This gauge dependence is inherited by the GSF, which transforms under a gauge transformation as,
\begin{equation}
 \tilde{F}^\mu - F^\mu = -\hh{g^{\mu\alpha}+u^\mu u^\alpha}\CD{u}^2\xi_\alpha-\tensor{R}{^\mu_\alpha_\beta_\gamma}u^\alpha\xi^\beta u^\gamma.
\end{equation}
For any practical calculation of the GSF, we must therefore choose a gauge to work in. A common choice in the self-force literature is the Lorenz gauge defined by,
\begin{equation}\label{eq:lorgauge}
	\CD{\alpha}\hh{h^{\alpha\mu}-\frac{1}{2}g^{\alpha\mu}g^{\beta\gamma}h_{\beta\gamma}}=0.
\end{equation}
However, in this work we work predominantly in the outgoing radiation gauge or ORG, which is defined by the conditions
\begin{align}
n^\mu h_{\mu\nu} &=0,\\
g^{\mu\nu}h_{\mu\nu} &=0.
\end{align}
These conditions can be met for vacuum perturbations. However, if a perturbation is sourced by some matter distribution, the ORG conditions cannot be met globally \cite{Ori:2002uv}. Trying to impose the ORG condition global on the perturbation produced by a point-particle results in a string-like (gauge) singularity extending from the particle to the horizon of the background geometry and/or infinity. This leads to various realization possibilities for the ORG \cite{Pound:2013faa}
\begin{itemize}
\item The \emph{half-string} gauges feature a half-string extending from the particle to either the background horizon or infinity. Elsewhere they are perfectly regular
\item The \emph{full-string} gauge has a sting extending from the background horizon to infinity through the particle.
\item The \emph{no-string} gauge is discontinuous along a hypersurface that includes the particle worldline and that separates the background horizon from infinity. On each side of this hypersurface the metric perturbation is realized as the regular half of one of the half-string gauges.
\end{itemize}

\subsection{l-mode regularization}\label{sec:lmodereg}
A key step in calculating the GSF is obtaining the regular metric perturbation through the subtraction
\begin{equation}\label{eq:hRdiff}
h_{\mu\nu}^\reg = h_{\mu\nu}^\ret - h_{\mu\nu}^\sing,
\end{equation}
(the derivatives of) which then need(s) to be evaluated on the particle worldline. This introduces a problem for any practical calculation since both $h_{\mu\nu}^\ret$ and  $h_{\mu\nu}^\sing$ are singular on the worldline (while their difference is not). We therefore need to introduce a regulator to allow for a systematic evaluation of the subtraction. In this work, we employ the $l$-mode regularization introduced by Barack and Ori \cite{Barack:2001gx,Barack:2001ph,Barack:2002bt}.

For any field $f(x)$ on the background spacetime the method defines its $l$-modes as,
\begin{equation}\label{eq:lmodedef}
f_l(x) \equiv \sum_{m=-l}^l
\Bh{\int_{S^2}\hspace{-8pt}\id{\Omega} f\bar{Y}_{lm}}Y_{lm}(z,\phi),
\end{equation}
where the integral is performed over a sphere of constant $t$ and $r$. The key property is that even if the field $f$ has a pole on some worldline, the $l$-modes remain finite (although possibly discontinuous).

The idea is then to evaluate \eqref{eq:GSFdef} independently on  $h_{\mu\nu}^\ret$ and $h_{\mu\nu}^\sing$, and calculate the $l$-modes of each. The subtraction can then be done at the level of the $l$-modes, and the sum of the resulting differences should produce a finite result for the GSF. However, as written Eq. \eqref{eq:GSFdef} is only defined on the worldline, and does not define the GSF as a field. In order to calculate the $l$-modes we therefore need to promote  \eqref{eq:GSFdef} to a field equation for $\mathcal{F}^\mu[h]$, or equivalently we need to extend the projector $P^{\mu\alpha\beta\gamma}$ in Eq.\eqref{eq:GSFprojector} to field off the worldline. This involves a (somewhat arbitrary) choice. Many of the details of the calculation (but not its final result) depend sensitively on this choice of \emph{extension}.

In this work, following \cite{Barack:2009ux}, we employ a ``rigid'' extension of $P^{\mu\alpha\beta\gamma}$, where it takes constant values on slices of constant $t$. 

With this choice of extension and adopting the Lorenz gauge it is possible to obtain a local Laurent expansion of $\mathcal{F}^\mu[h^\sing]$, and subsequently the large $l$ behaviour of its $l$-modes \cite{Barack:2002bt,Barack:2009ux,Mino:2001mq},
\begin{equation}\label{eq:ABCdef}
\begin{split}
F^{\mu,\pm}_{\sing,l} &\equiv\lim_{x\to x_0^\pm} \Fext^\mu_{\sing,l}\\
 &=  \pm L A^{\mu}_\Lor + B^{\mu}_\Lor +\frac{C^{\mu}_\Lor}{L} + \bigO(L^{-2}), 
\end{split}
\end{equation}
with $L:=l+1/2$, and the $\pm$ sign depends on the radial direction from which $x_0$ is approached. Furthermore, one can show that,
\begin{equation}\label{eq:Ddef}
D^{\mu}_\Lor \equiv \sum_l F^{\mu,\pm}_{l,\sing} \mp L A^{\mu}_\Lor - B^{\mu}_\Lor -\frac{C^{\mu}_\Lor}{L}=0.
\end{equation}
Consequently, if one can obtain the $l$-modes $F^{\mu}_{l,\Lor}$ of the retarded field with same choice of gauge and extension, then one can obtain the Lorenz gauge GSF using the mode-sum formula,
\begin{equation}\label{eq:modesum}
F^\mu_\Lor = \Bh{\sum_l F^{\mu,\pm}_{l,\Lor} \mp L A^{\mu}_\Lor - B^{\mu}_\Lor -\frac{C^{\mu}_\Lor}{L}}-D^{\mu}_\Lor.
\end{equation}
The quantities $A^{\mu}_\Lor$, $B^{\mu}_\Lor$, $C^{\mu}_\Lor$, and $D^{\mu}_\Lor$ are collectively known as \emph{regularization parameters}.

However, in this work we obtain the retarded metric perturbations not in the Lorenz gauge, but in the outgoing radiation gauge. Calculation of the GSF in radiation gauges was studied by Pound, Merlin, and Barack in \cite{Pound:2013faa}. They concluded that one can calculate the GSF in the `half-string' gauges using the mode-sum formula provided that the limit towards the particle is taken from the regular side. In this case, the $A$, $B$, and $C$ parameter are identical to the Lorenz gauge ones, provided one uses the same extension. The $D$ parameter, however, acquires a non-zero correction which is hard to calculate in practice. It is also possible to calculate the GSF in the no-string radiation gauge. In this case it is necessary to take the limit towards the particle from both sides and average the result. It turns out that with that prescription all regularization parameters (including $D$) take their Lorenz gauge values. The no-string radiation gauge mode-sum formula is thus given by
\begin{equation}\label{eq:modesumavg}
F^\mu_\Rad = \Bh{\sum_l \frac{F^{\mu,+}_{l,\Rad}+F^{\mu,-}_{l,\Rad}}{2} - B^{\mu}_\Lor -\frac{C^{\mu}_\Lor}{L}}-D^{\mu}_\Lor.
\end{equation} 

\subsection{Radiation gauge metric reconstruction}
In this work, we avoid the difficulties of directly solving the linearized Einstein equation on a Kerr background, by trying to recover the metric perturbation from the Weyl scalar $\psi_4$, which can be obtained for particles on generic bound orbits in Kerr spacetime by solving the spin-(-2) Teukolsky equation in the frequency domain \cite{Drasco:2005kz,Fujita:2009us}. That this should be possible was first hinted at by Wald \cite{Wald:1973}, who showed that $\psi_4$ contains all information about the metric perturbation modulo a perturbation within the Kerr family of solutions and gauge information.

The first steps towards this goal were set by Chrzanowski, Cohen, Kegeles, and Wald \cite{Chrzanowski:1975wv,Cohen:1974cm,Kegeles:1979an,Wald:1978vm}, who showed that given a solution of the vacuum spin-($\pm2$) Teukolsky equation one can obtain a vacuum solution of the linearized Einstein equation. The operators that achieves this are essentially the adjoint of the operators that construct the sources for $\psi_0$ and $\psi_4$ from the energy-momentum tensor \cite{Wald:1978vm}. However, if one calculates $\psi_4$ from the metric perturbation obtained from a vacuum solution spin-(-2) Teukolsky equation, one does not recover the same vacuum solution of the spin-(-2) Teukolsky equation.  The vacuum solutions of the Teukolsky equation are therefore not the Weyl scalars $\psi_0$ and $\psi_4$. Instead they are different fields known as Hertz potentials.

The problem of obtaining the Hertz potential corresponding to a certain vacuum solution of $\psi_0$ and $\psi_4$ involves inverting a fourth-order differential equation \cite{Lousto:2002em}. This was first tackled by Ori \cite{Ori:2002uv}, who showed how to obtain the spin-(+2) Hertz potential corresponding to a $\psi_0$ by inverting the differential equation mode-by-mode in the frequency domain. A similar procedure was employed by Keidl et al. \cite{Keidl:2010pm} to obtain the spin-(-2) Hertz potential from $\psi_0$. In a previous paper \cite{vandeMeent:2015lxa} the author showed how to obtain the spin-(+2) Hertz potential from $\psi_4$. We will use this last procedure which results in a metric perturbation in the ORG.

As explained in previous papers \cite{vandeMeent:2015lxa,vandeMeent:2016hel}, the metric perturbation produced by a point particle on an eccentric orbit, for which the frequency domain source will have support over a finite range in the radial direction, can be obtained by solving the Teukolsky equation for $\psi_4$, executing the inversion and metric reconstruction in the vacuum regions away from the source, and analytically extending those vacuum metric perturbations back to the particle worldline. This `extended homogeneous solutions' procedure naturally produces a metric in the `no-string' outgoing radiation gauge, and works without alteration for inclined orbits.

The final step is to complete the metric by finding the missing perturbations within the Kerr family. Since, the no-string solution is discontinuous we need to find separate perturbations in each half. In \cite{vandeMeent:2017fqk}, it was shown how these can be recovered for general sources. In particular, for a point particle on a generic orbit the Kerr perturbations vanish on the inner half of the solution, while on the outer half they are given by
\begin{equation}\label{eq:completion}
h^{\mathrm{comp},+}_{\mu\nu} = \nE \pd{g_{\mu\nu}}{M}\Biggr|_{J}+
\nL \pd{g_{\mu\nu}}{J}\Biggr|_{M},
\end{equation}
where $J=M a$ is the angular momentum of the Kerr metric, 

\section{Method}\label{sec:method}
Our method for calculating the GSF on generic Kerr geodesics is in many respects identical  to the methods used for calculating the regular metric and GSF on equatorial eccentric orbits described in \cite{vandeMeent:2015lxa} and \cite{vandeMeent:2016hel}. In this section we will therefore give only a brief outline of these methods and focus on the details that are different in the generic case.

\subsection{Weyl scalar \texorpdfstring{$\psi_4$}{psi4} and Hertz potential}
As before in \cite{vandeMeent:2015lxa,vandeMeent:2016hel}, we use the formalism of Mano, Suzuki, and Takasugi (MST) \cite{Mano:1996gn,Mano:1996vt} to solve the (homogeneous) Teukolsky equation, largely following the numerical implementation of Fujita and Tagoshi \cite{Fujita:2004rb,Fujita:2009us}. The method of variation of parameters can then be used to find the $\psi_4$ generated by a particle of a generic Kerr geodesic, as first demonstrated by Drasco and Hughes \cite{Drasco:2005kz}. Details of our arbitrary precision numerical implementation are forthcoming \cite{Meent:2015a}.

Once $\psi_4$ is known, we can use the procedure described in \cite{vandeMeent:2015lxa} to obtain the corresponding spin-(+2) Hertz potential $\Psi_{+2}^\pm$ in the asymptotic vacuum regions toward infinity (``$+$'') and  towards the horizon  (``$-$''). These are then analytically extend towards the particle. The result has the form,
\begin{equation}\label{eq:PsiS}
\begin{split}
\Psi_{+2}^\pm = \frac{1}{\sqrt{2\pi}}\sum_{\spl m\omega} \Psi_{\spl m\omega}^\pm \R[\pm]{2}{\spl m\omega}(r)\SWSH{2}{\spl m\omega}(z)\ee^{\ii m\phi-\ii\omega t},
\end{split}
\end{equation}
where $\Psi_{\spl m\omega}^\pm$ are the mode amplitudes obtained through solving the inhomogeneous Teukolsky equation and the inversion procedure. The $\R[\pm]{2}{\spl m\omega}(r)$ are homogeneous solutions of the spin-(+2) Teukolsky equation with outgoing boundary conditions at either infinity or the horizon. The $\SWSH{2}{\spl m\omega}(z)$ are spin-weighted spheroidal harmonics with spin weight $+2$. Finally, the discrete $\omega$-sum is over the set $\{m\Omega_\phi+k\Omega_z+n\Omega_r|m,k,n\in\ZZ\}$, where the $\Omega_i$ are the Boyer-Lindquist coordinate time frequencies.

\subsection{GSF coefficients}\label{sec:GSFcoef}
The expression in \eqref{eq:PsiS} can be used in the procedure described in \cite{vandeMeent:2016hel} to obtain the GSF. The steps are
\begin{enumerate}
\item Apply the ORG metric reconstruction operator.
\item Apply $P^{\mu\alpha\beta\gamma}\CD{\alpha}$ to obtain $\Fext_{\Rad}^{\mu,\pIH}$, the field extended form of the GSF.
\item Use
\begin{equation}
\SWSH{s}{\spl m\omega}(z)=\sum_{l}\Sb{s}{m\omega}{\spl}{l} \Y{s}{lm}(z),
\end{equation}
where the $\Sb{s}{m\omega}{\spl}{l}$ are obtained through the method of \cite{Hughes:1999bq} for expanding  spin-weighted spheroidal harmonics in spin-weighted spherical harmonics.
\item Use
\begin{equation}
\SWL{s}=\sqrt{1-z^2}\hh{\partial_z+\frac{\ii}{1-z^2}\partial_\phi-\frac{s z}{1-z^2}},
\end{equation}
to eliminate any $z$ derivatives in favour of spin-lowering operators.
\item Re-expand the resulting variety of spin-weighted spherical harmonics to regular spherical harmonics, using
\begin{align}
\Y{2}{l_1m}(z)&=\sum_{l_2}
\frac{
\YA{2}{m}{l_1}{l_2} \Y{}{l_2m}(z)
}{
1-z^2
}
,\label{eq:YA1}\\
\Y{1}{l_1m}(z)&=\sum_{l_2}
\frac{
\YA{1}{m}{l_1}{l_2}\Y{}{l_2m}(z)
}{
\sqrt{(l_1-1)(l_1+2)}\sqrt{1-z^2}
}
,\\
\Y{0}{l_1m}(z)&=\sum_{l_2}\frac{
\YA{0}{m}{l_1}{l_2}\Y{}{l_2m}(z)
}{
\sqrt{(l_1-1)l_1(l_1+1)(l_1+2)}
} ,\\
\Y{-1}{l_1m}(z)&=\sum_{l_2}\frac{
\YA{-1}{m}{l_1}{l_2}
\Y{}{l_2m}(z)
\sqrt{(l_1-2)!}
}{
\sqrt{(l_1+2)!l_1(l_1+1)}\sqrt{1-z^2}
} ,\label{eq:YA4}
\end{align}
where the $\YA{s}{m}{l_1}{l_2}$ are defined in \cite{vandeMeent:2016hel}.
\end{enumerate}
The result is an expression of the form,
\begin{align}\label{eq:GSF2}
\Fext_{\Rad}^{\mu,\pIH} &=\begin{aligned}[t]
  \sum_{\substack{m{\omega}si\\l_1l_2\spl}} 
&\MC^\mu_{m{\omega}si}(r,z)\Psi_{\spl m\omega}^\pIH
\R[\pIH,(i)]{2}{\spl m\omega}(r)
\Sb{2}{m\omega}{\spl}{l_1}
\\
&\times
\YA{s}{m}{l_1}{l_2}
\Y{}{l_2m}(z){\ee}^{\ii m\phi-\ii\omega t}+c.c.,&
\end{aligned}
\end{align}
where the $\MC^\mu_{m{\omega}si}(r,z)$ are coefficient functions determined through the procedure above. At this point the procedure starts to diverge from the equatorial case. In the equatorial case we could use the up/down symmetry of the source to resolve the ``+c.c.'' terms in a simple form. This symmetry is no longer available for generic orbits (which only satisfy up/down symmetry on average). As a result, we will just leave the ``+c.c.'' terms as they are.

The form of \eqref{eq:GSF2} is almost that of an expansion in $l$-modes as needed for our mode-sum regularization. However, as it stands the $\MC^\mu_{m{\omega}si}(r,z)$ still depend on the field coordinate $z$. To remedy this situation we replace $\MC^\mu_{m{\omega}si}(r,z)$ by its Taylor expansion around the polar position of the particle $z_0$. Truncating this expansion amounts to changing the field extension of the self-force. As mentioned in Sec. \ref{sec:lmodereg}, the values of the regularization parameters depends on the extension. In order to ensure that the values are unchanged we need that the  extension agrees on the first three terms of the Taylor expansion. Hence we keep the first three terms of the Taylor expansion of $\MC^\mu_{m{\omega}si}(r,z)$.

Expanding the result we can eliminate terms of the form $z^n \Y{}{l_2m}(z)$ using the re-expansion 
\begin{equation}
z^n\Y{}{l_1m}(z)=\sum_{l_2}\YB{n}{m}{l_1}{l_2}
 \Y{}{l_2m}(z),
\end{equation}
where
\begin{equation}
\begin{split}
\YB{1}{m}{l_1}{l_2} = (-1)^{m+l_1+1}&(l_1-l_2)
\\
\times&\sqrt{\frac{l_1+l_2+1}{2}}
\begin{pmatrix}
1 & l_1 & l_2\\
0 & m	& -m
\end{pmatrix},
\end{split}
\end{equation}
and
\begin{equation}
\YB{n+1}{m}{l_1}{l_2}=\sum_\ell\YB{1}{m}{l_1}{\ell}\YB{n}{m}{\ell}{l_2}.
\end{equation}

From the resulting expression we obtain the $l$-modes of GSF,

\begin{align}\label{eq:GSFlmodes}
F_{\Rad,l}^{\mu,\pIH} &=\begin{aligned}[t]
  \sum_{\substack{m{\omega}sin\\l_1l_2\spl}} 
&\MC^\mu_{m{\omega}sin}(r_0,z_0)\Psi_{\spl m\omega}^\pIH
\R[\pIH,(i)]{2}{\spl m\omega}(r_0)
\\
&\hspace{-2em}\times
\Sb{2}{m\omega}{\spl}{l_1}
\YA{s}{m}{l_1}{l_2}
\YB{n}{m}{l_2}{l}
{\ee}^{\ii m\phi_0-\ii\omega t_0}+c.c.,&
\end{aligned}
\end{align}
where the $\MC^\mu_{m{\omega}sin}(r_0,z_0)$ are a new set of coefficients that now only depend on the particle orbit. Although the formal expression given here is not much different than the one in \cite{vandeMeent:2016hel}, the full explicit expression is significantly more complicated. This can be expressed in terms of the leaf count of the \emph{Mathematica} representations of the $\MC^\mu_{m{\omega}sin}(r_0,z_0)$. For equatorial orbits the leaf count of these expressions was less than 200,000. For the new expressions for generic orbits the leaf count is nearly 6 million.

As a final remark in this section we note that using \eqref{eq:t}, \eqref{eq:phi}, and $\omega = m\Omega_\phi+k\Omega_z+n\Omega_r$, we can rewrite
\begin{align}
{\ee}^{\ii m\phi_0-\ii\omega t_0} &= {\ee}^{\ii (m \phi_z-k q_z -\omega t_z)}{\ee}^{\ii (m \phi_t-n q_r -\omega t_r)}.
\end{align}
Consequently, the l-modes of the GSF can be expressed as functions of $(a,p,e,\zmax)$ and $(q_r,q_z)$. As expected, we can express the orbital variation of the GSF purely in terms of the $(q_r,q_z)$-torus.

\subsection{Mode-sum and completion}
Once we have obtained the $l$-modes we can subtract the regularization parameters calculated in \cite{Barack:2002bt,Barack:2009ux} and calculate the mode-sum \eqref{eq:modesumavg} to obtain (the reconstructed) piece of the GSF. We follow the procedure outlined in \cite{vandeMeent:2015lxa} to numerically fit the large $l$-tail of the sum to accelerate convergence of the sum. This procedure is performed separately for each $(q_r,q_z)$ point along the orbit. To obtain the full GSF we need to add the piece coming from the Kerr-type perturbations of the background. As shown in \cite{vandeMeent:2017fqk}, this piece is given by Eq. \eqref{eq:completion}. The contribution to the GSF is found by simply applying \eqref{eq:GSFdef}.

The GSF obtained in this manner contains all gauge invariant information contained in the GSF. However, many quantities that we like to calculate and compare between different calculations such as the Detweiler redshift \cite{Detweiler:2008ft} or the periapsis precession \cite{vandeMeent:2016pee}, are only invariant under the restricted class of gauge transformations that remain small over the inspiral timescale $\sim \mr^{-1}$. To calculate such quasi-invariants one would need to fix the remaining gauge freedom, adding a gauge correction to the completion inside the orbit \cite{gaugecompletion}. Presumably, such a correction is also needed to evolve inspirals. However, in this work we do not add such corrections as they are not needed here.

\section{Results}\label{sec:results}
\begin{figure}
	\includegraphics[width=\columnwidth]{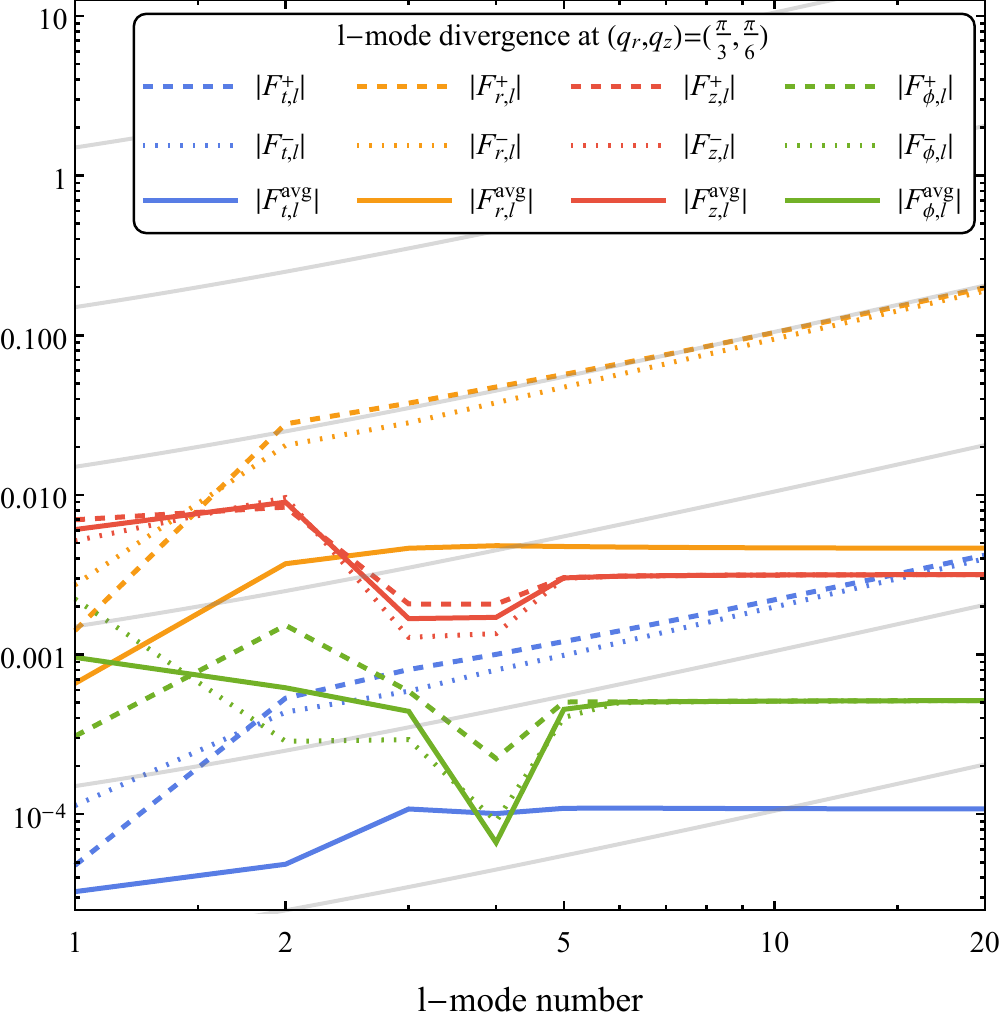}
	\caption{The $l$-modes of the various components of the GSF on a geodesic with $(a,p,e,\zmax) = (0.9,10,0.1,0.1)$, shown at the point along the orbit identified by $(q_r,q_z)= (\pi/3,\pi/6)$. This point is indicative of the generic behaviour (certain special points such as periapsis and apapsis will show better convergence behaviour). The grey lines are reference lines of $L=l+1/2$. As expected the $\pm$ parts of the $t$ and $r$ components diverge with $L$. The parameters $A_z$ and $A_\phi$ vanish \cite{Barack:2009ux,Barack:2002mh}. Consequently, we see $\pm$ parts of the $r$ and $\phi$ l-modes converge to a constant, just like all the two-side average parts.
	}\label{fig:lmodediv}
\end{figure}
We have implemented the above method for calculating the GSF on generic Kerr orbits in our arbitrary precision \emph{Mathematica} code. This implementation is considerably more computation intensive than the implementation for eccentric orbits. There are three main contributing factors
\begin{enumerate}
\item As mentioned above, the various expressions for forming the GSF are considerably more complex, taking more time to evaluate and consuming more memory.
\item For generic orbits we now have a 2-dimensional spectrum of frequency modes for each $(l,m)$-mode. As a result we need to compute many more modes for a single orbit.
\item Orbits are now parametrized by two independent phases. Consequently, we need to sample the orbits at many more points. 
\end{enumerate}
As a result where moderately eccentric equatorial orbits would require at most tens of CPU hours to calculate the GSF, to calculate the GSF on a single inclined Kerr geodesic with modest eccentricity requires up to $10^4$ CPU hours. Luckily the large number of modes, means that the code is embarrassingly parallelizable, easily running on 400+ cores at an usage efficiency upwards of 90\%.

Therefore, for this work we have chosen to limit ourselves to a limited number of 5 orbits with fixed spin (${a=0.9}$), semilatus rectum (${p=10}$), and eccentricity (${e=0.1}$), while varying the inclination from ${\zmax=0.1}$ to ${\zmax=0.9}$. In the following sections we first present some consistency checks on our results. We then provide some graphical representation of the GSF results.

\subsection{Consistency checks}
\subsubsection{Regularization parameters}
\begin{figure}
\includegraphics[width=\columnwidth]{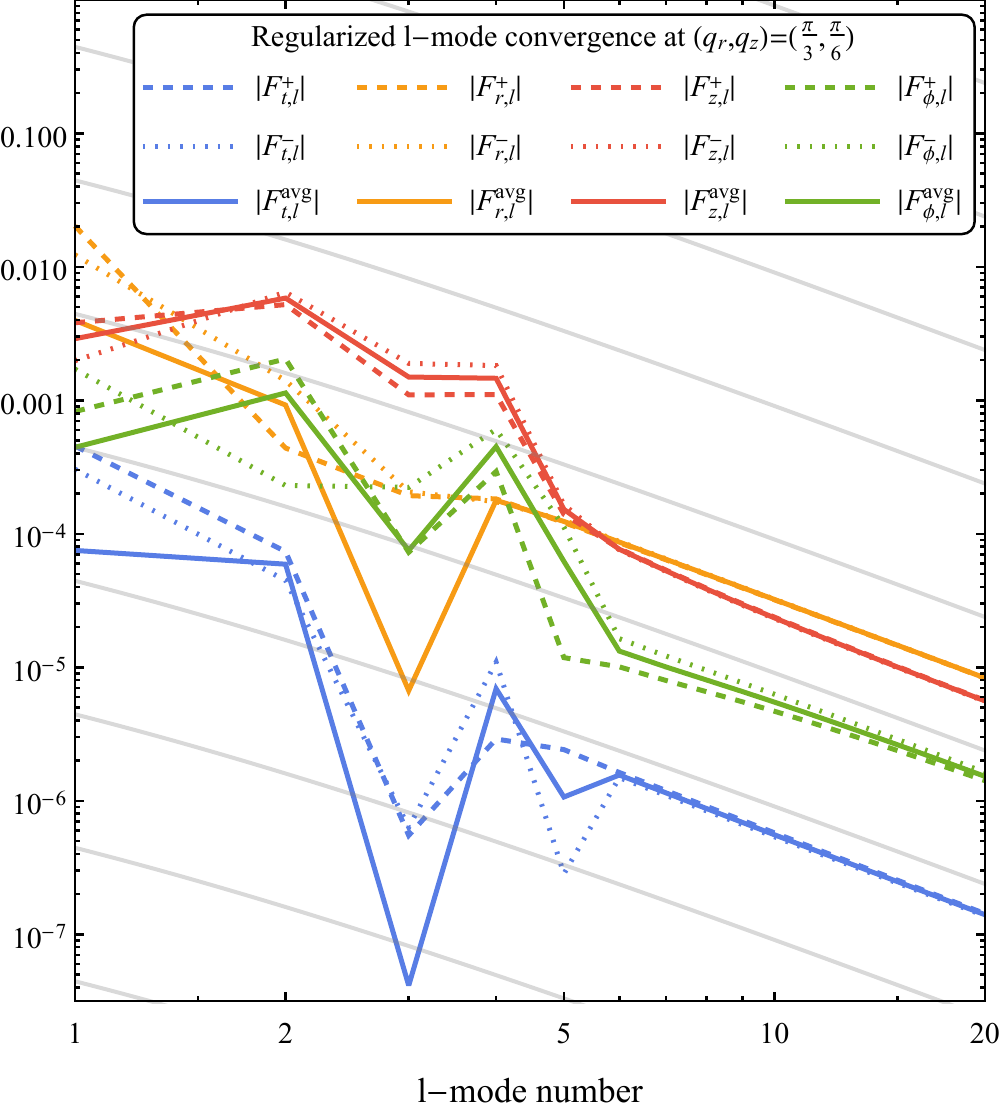}
\caption{The same $l$ modes as in Fig. \ref{fig:lmodediv} after subtracting the Lorenz gauge regularization parameters. At large $l$, all components of the GSF conform with $L^{-2}$ behaviour indicated by the grey reference lines. This is a stringent check on the validity of our method and numerical implementation.
}\label{fig:lmodeconv}
\end{figure}
\input{balancelaw.tab}
\begin{figure*}[t]
\includegraphics[width=\textwidth]{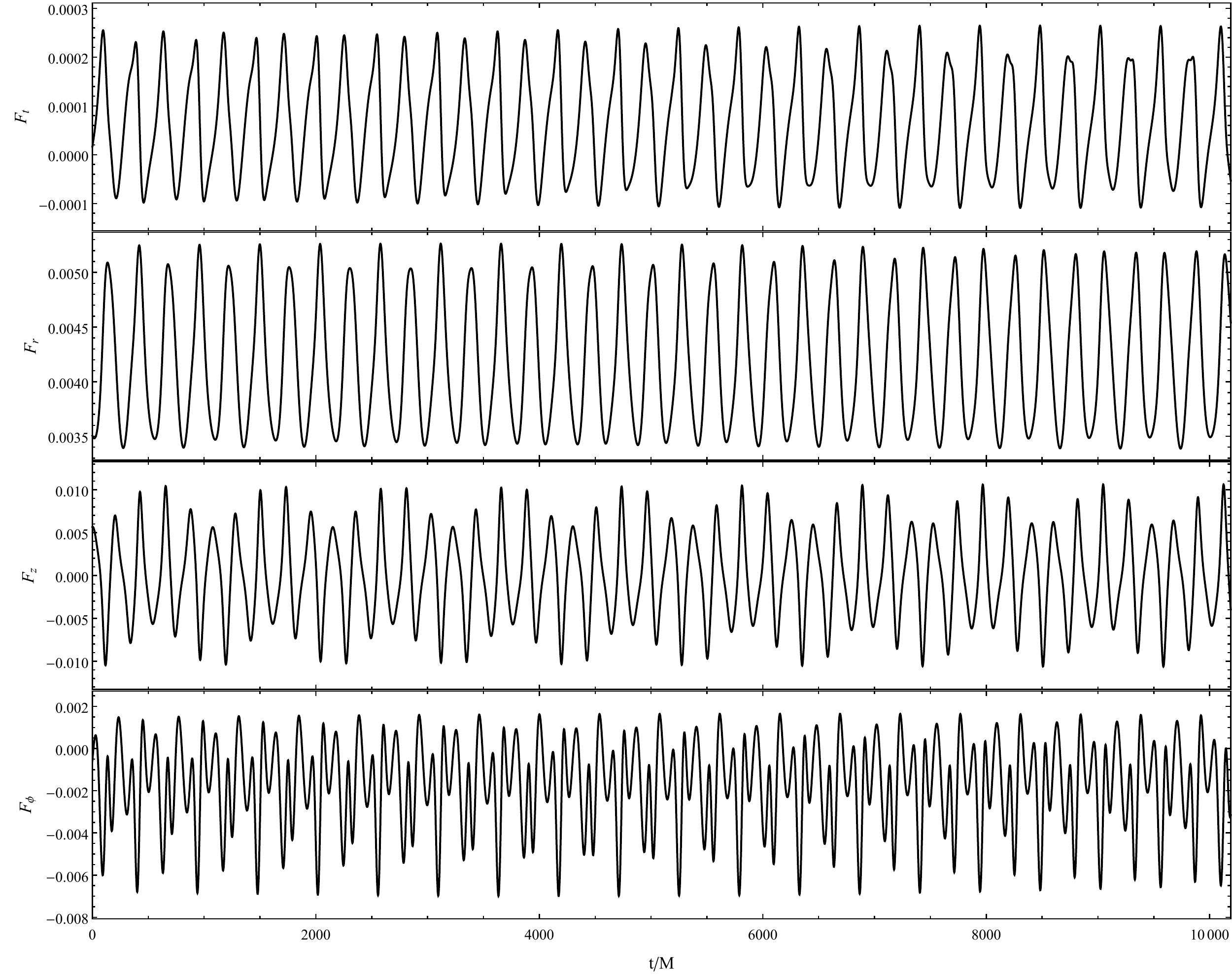}
\caption{Time series data of the GSF on an orbits with $(a,p,e,\zmax)=(0.9,10,0.1,0.5)$. Because of the bi-periodic nature of the GSF none of the modulation patterns ever repeat.
}\label{fig:timeseries}
\end{figure*}
A key consistency check for any self-force calculation is comparison of the large $l$ behaviour of the $l$-modes with the analytically calculated regularization parameters. After subtracting the regularization parameters from the l-modes, the remainder should decay with $(L=l+1/2)^{-2}$ for large $l$. This requires a large degree of cancellation between the two calculations. As a result, it is virtually impossible to make a mistake in either calculation without creating disagreement between the results.

We start by plotting in Fig. \ref{fig:lmodediv} the $l$-modes of the GSF on a Kerr geodesic with $(a,p,e,\zmax) = (0.9,10,0.1,0.1)$ evaluated at the point $(q_r,q_z)=(\pi/3,\pi/6)$. This point has been chosen as suitably representative of a generic point along the orbit. For certain special points along the orbit such as the turn points, the $l$-modes will decay without the need for regularization. We will avoid these points for this test. As can be seen in Fig. \ref{fig:lmodediv}, the $l$-modes of the ``outside'' ($+$) and ``inside'' ($-$) values of the $t$ and $\phi$ components of the GSF grow linearly with $L$, as is expected since these components have non-zero $A$ parameters. The parameters $A_z$ and $A_\phi$ vanish \cite{Barack:2002mh}, and indeed we see that the ``outside'' ($+$) and ``inside'' ($-$) values of the corresponding $l$-modes converge to a constant for large $l$. Similarly, as required by Eq. \eqref{eq:modesumavg}, the $l$-modes of the two-sided averages of all components also converge to a constant.

Fig. \ref{fig:lmodeconv} shows the same modes as in Fig. \ref{fig:lmodediv} after subtracting the analytical regularization parameters from \cite{Barack:2009ux}. As we should expect, all $l$-modes exhibit a $L^{-2}$ decay at large $l$. As mentioned above this provides an extremely stringent test on the validity of our methods and implementation. In addition it provides independent confirmation of the results of \cite{Pound:2013faa} showing that the Lorenz gauge $A$, $B$, and $C$ regularization parameters can be used to regularize the GSF obtained in half-string or no-string realizations of the radiation gauge. Finally, this provides the first numerical verification of the analytical calculation of the regularization parameters for generic Kerr geodesics \cite{Barack:2002bt,Barack:2009ux}. 

\subsubsection{Flux balance law}

A second important verification test of our results, is checking whether the ``flux balance law'' is satisfied. The ``flux balance law''  \cite{Mino:2003yg,Mino:2005an,Mino:2005yw,Sago:2005fn,Sago:2005fn} says that the orbit averaged changes to the constants of motion $\nE$, $\nL$, and $\nQ$ due to the local GSF should match the change to this constants of motion inferred from the gravitational wave flux at infinity and the background horizon.

The local changes to the constants of motion can be calculated as follows
\begin{align}
\avg{\d{\nE^\mathrm{GSF}}{\pt}} &=- \frac{\mr\avg{\d{\mt}{\pt}}}{(2\pi)^2} \int_{-\pi}^{\pi}  \int_{-\pi}^{\pi} F_t \id{q_r}\id{q_z},\\
\avg{\d{\nL^\mathrm{GSF}}{\pt}} &= \frac{\mr\avg{\d{\mt}{\pt}}}{(2\pi)^2} \int_{-\pi}^{\pi}  \int_{-\pi}^{\pi} F_\phi \id{q_r}\id{q_z} \text{, and}\\
\avg{\d{\nQ^\mathrm{GSF}}{\pt}} &= \frac{\mr\avg{\d{\mt}{\pt}}}{(2\pi)^2} \int_{-\pi}^{\pi}  \int_{-\pi}^{\pi}2u^\alpha K_{\alpha\beta} F^\beta\\
&\qquad\qquad-2(\nL-a\nE)(F_\phi+aF_t)\id{q_r}\id{q_z} 
.\notag
\end{align}
Teukolsky and Press \cite{Teukolsky:1973ha,Teukolsky:1974yv} already showed how the average fluxes of $\nE$ and $\nL$ can be extracted from the behaviour of $\psi_4$ at the horizon and infinity. A similar result for $\nQ$ was obtained in 2005 by Sago et al. \cite{Sago:2005fn,Sago:2005gd}.

In table \ref{tab:blaw} we compare the results of the flux calculations to the local averages obtained from the GSF on our five test orbits. The results agree to all available digits. This provides another verification of our implementation and methods. At the same time, it provides the first direct numerical test of the formula derived by Sago et~al.~\cite{Sago:2005fn,Sago:2005gd} for obtaining the average change of the Carter constant from the flux.

\subsection{Sample results}
\subsubsection{Time series}

In Fig. \ref{fig:timeseries} we plot the components of the GSF on a geodesic with $(a,p,e,\zmax)=(0.9,10,0.1,0.5)$ as a function of coordinate time. These time series display the bi-periodic  nature of the GSF on a generic Kerr geodesic, showing imprints of components of the radial frequency $\Omega_r = 0.02325$ and of the polar frequency $\Omega_z = 0.0291866$. Since these frequencies are incommensurate the oscillation patterns never really repeat. Most of the modes are dominated by oscillations compatible with the $\Omega_r$ and $\Omega_z$ frequencies. The notable exception is the $\phi$ components which oscillates on a shorter timescale, we will comeback to the cause of this in the next section. 

\subsubsection{Torus plots}
\begin{figure*}[!t]
\includegraphics[width=\textwidth]{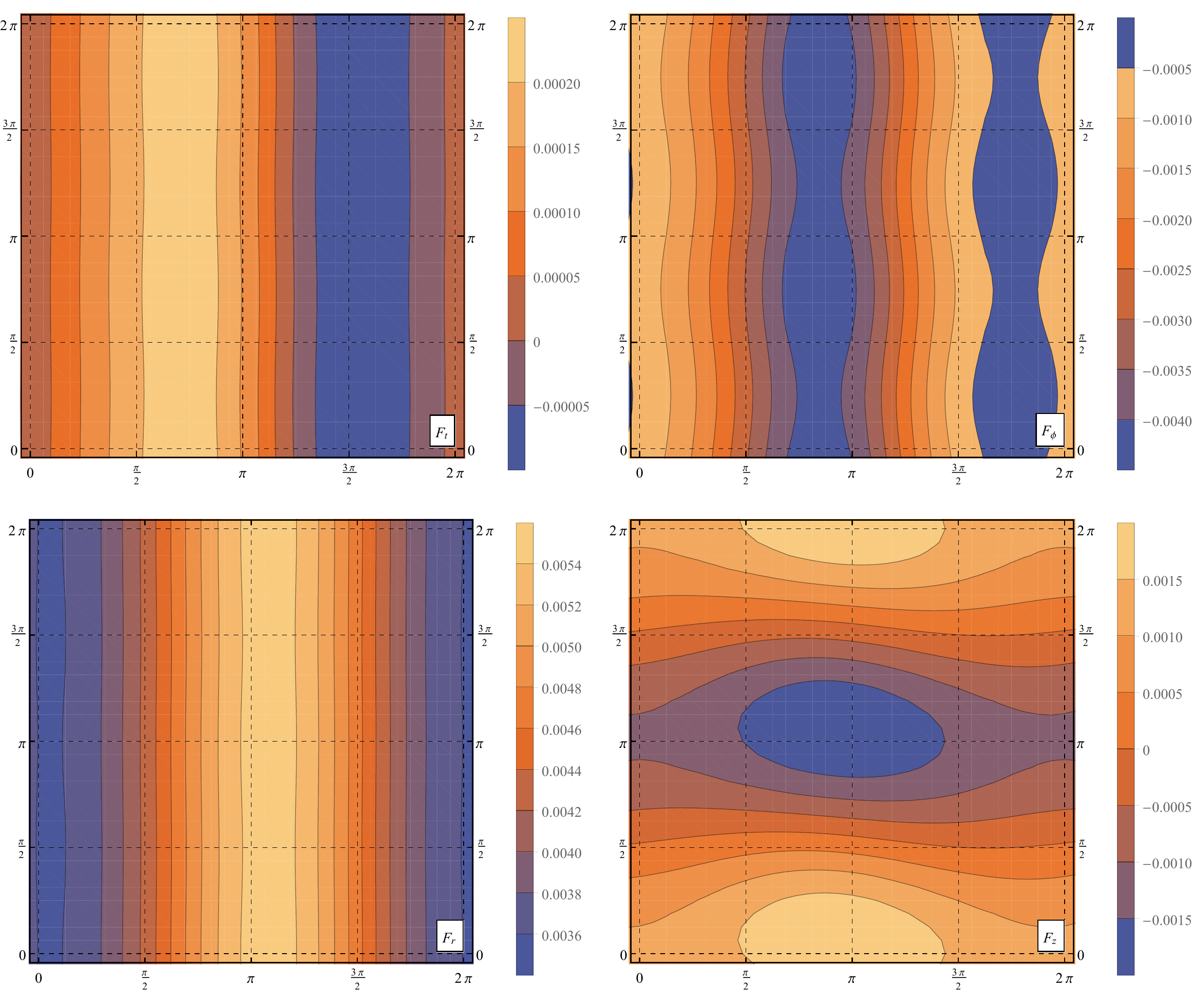}
\caption{GSF as a function on the torus for an orbit with $(a,p,e,\zmax)=(0.9,10,0.1,0.1)$. The horizontal axis displays changing $q_r$, while the vertical axis displays $q_z$.
}\label{fig:forceplotz1}
\end{figure*}
\begin{figure*}[!t]
\includegraphics[width=\textwidth]{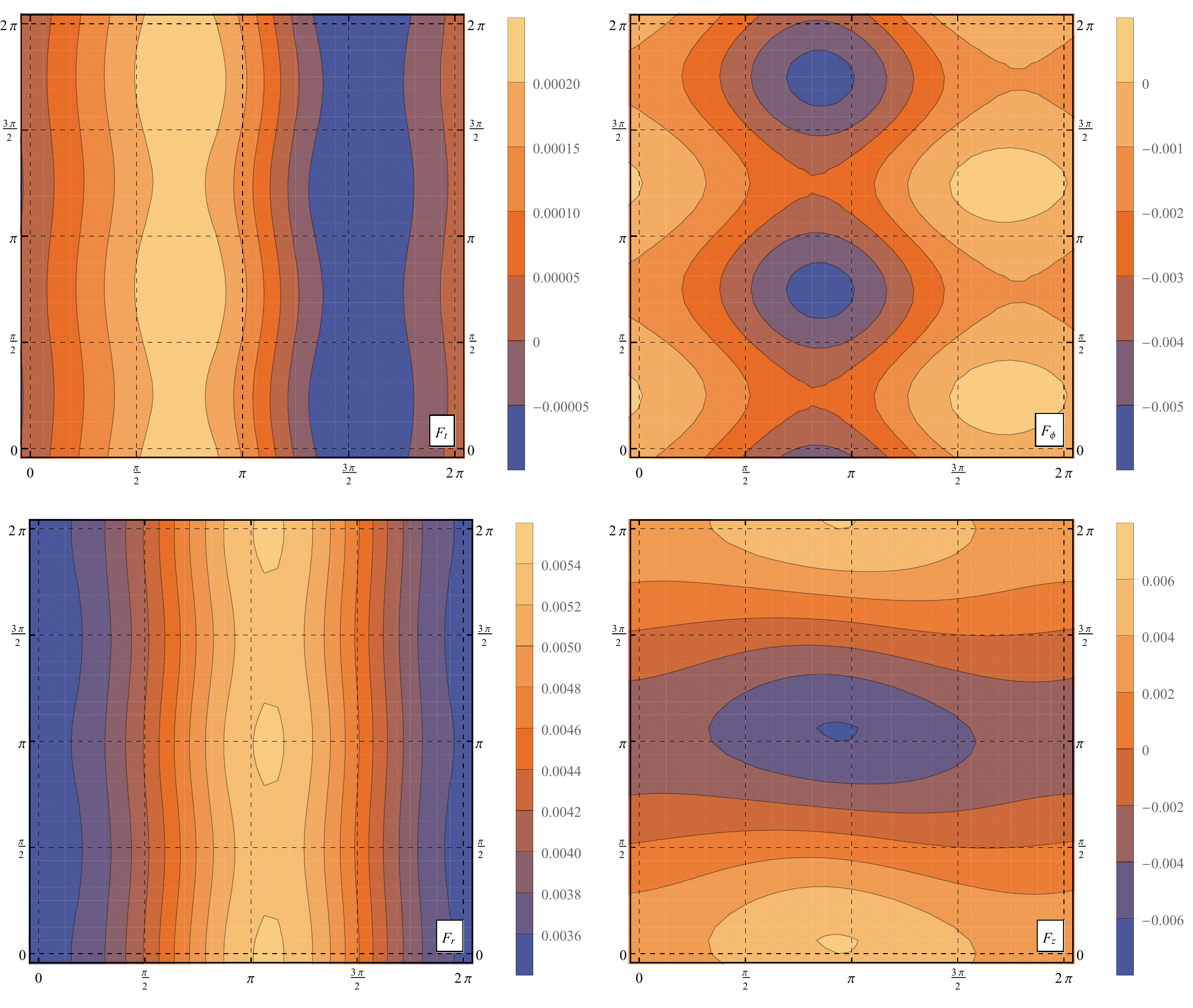}
\caption{GSF as a function on the torus for an orbit with $(a,p,e,\zmax)=(0.9,10,0.1,0.3)$. The horizontal axis displays changing $q_r$, while the vertical axis displays $q_z$.
}\label{fig:forceplotz31}
\end{figure*}
\begin{figure*}[!t]
\includegraphics[width=\textwidth]{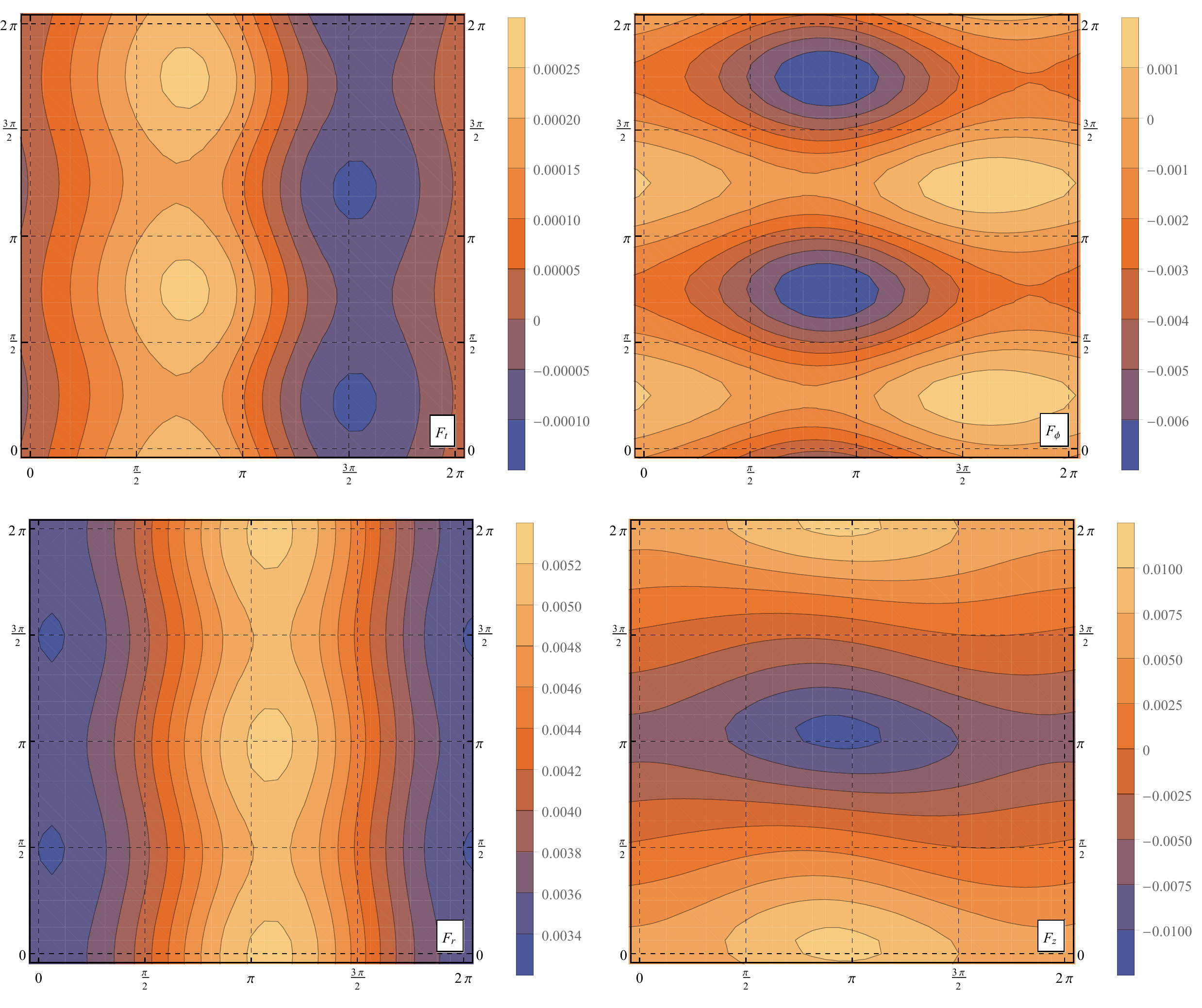}
\caption{GSF as a function on the torus for an orbit with $(a,p,e,\zmax)=(0.9,10,0.1,0.5)$. The horizontal axis displays changing $q_r$, while the vertical axis displays $q_z$.
}\label{fig:forceplotz5}
\end{figure*}
\begin{figure*}[!t]
	\includegraphics[width=\textwidth]{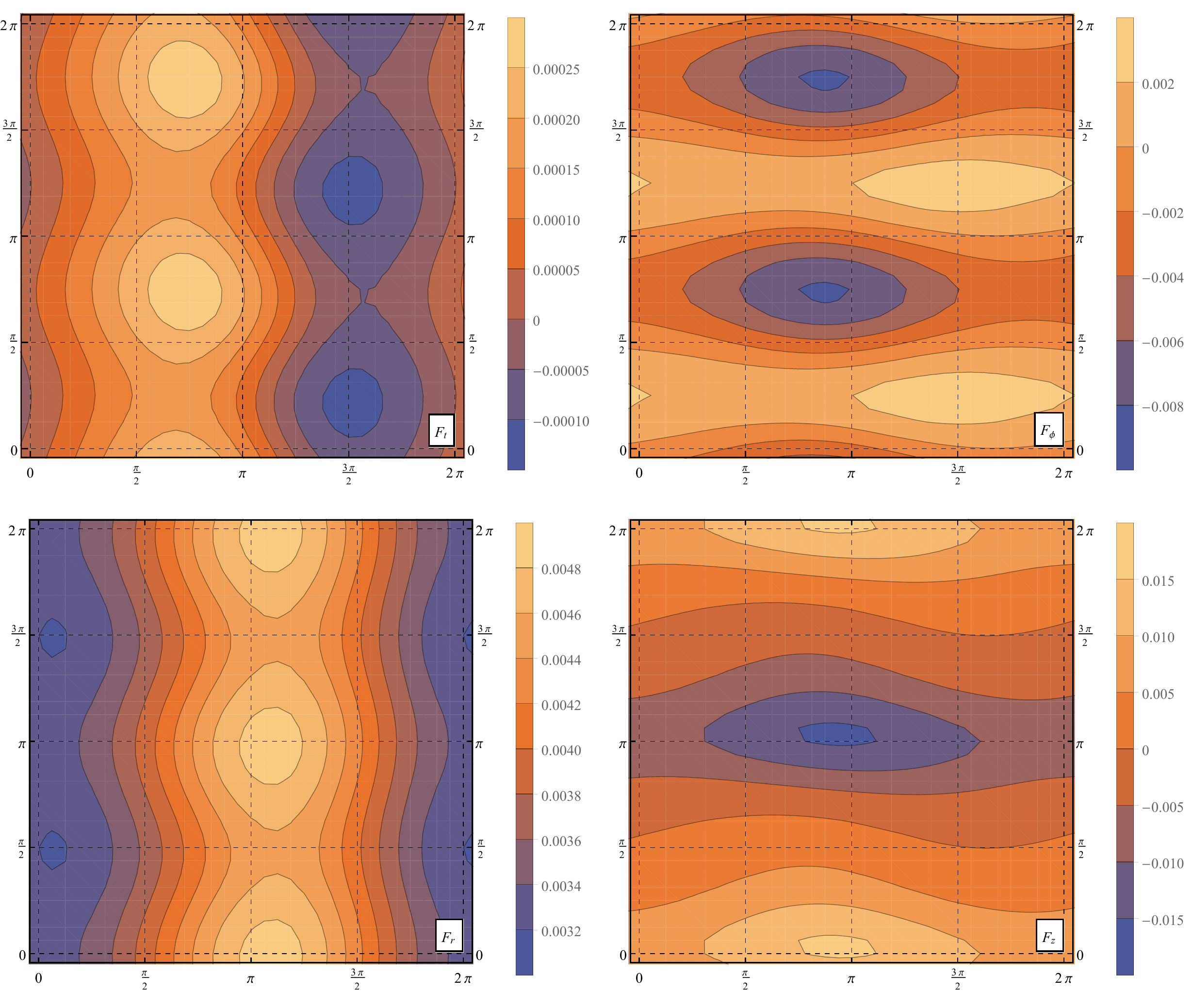}
	\caption{GSF as a function on the torus for an orbit with $(a,p,e,\zmax)=(0.9,10,0.1,0.7)$. The horizontal axis displays changing $q_r$, while the vertical axis displays $q_z$.
	}\label{fig:forceplotz7}
\end{figure*}
\begin{figure*}[!t]
	\includegraphics[width=\textwidth]{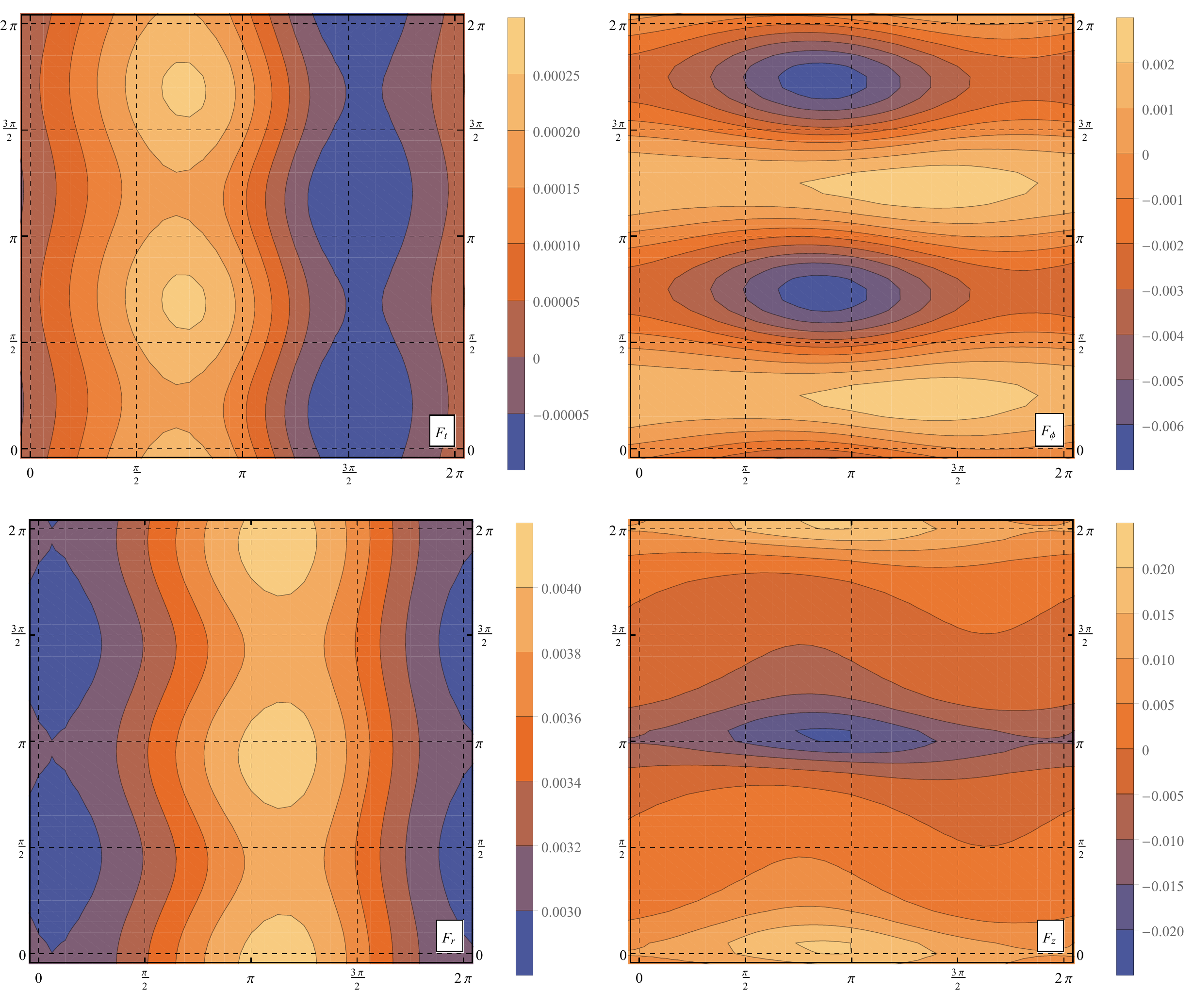}
	\caption{GSF as a function on the torus for an orbit with $(a,p,e,\zmax)=(0.9,10,0.1,0.9)$. The horizontal axis displays changing $q_r$, while the vertical axis displays $q_z$.
	}\label{fig:forceplotz9}
\end{figure*}

Although intuitive to read, the time series plots are not very informative about the features of the generic Kerr geodesic self-forces. In this section, we take a more integral approach to plotting the GSF on generic geodesics. As noted at the end of Sec. \ref{sec:GSFcoef}, the GSF on generic geodesic reduces to a pure function of the orbital phases. This is, of course, a natural consequence of the axisymmetry and stationarity of the background. Consequently, it makes sense to plot the components of the GSF as a function of the torus coordinates $(q_r, q_z)$. We do so in Fig. \ref{fig:forceplotz1}-\ref{fig:forceplotz9}. 

These figures are bit less intuitive to read, but do contain all available information about the GSF on a generic orbit. To help read the plots, note that the horizontal axis plots $q_r$. Consequently, the vertical lines at $q_r=0$ and $q_r=2\pi$ correspond to apapsis passages of the radial motion, and the vertical line at $q_r=\pi$ corresponds to periapsis passages. Similarly, since the vertical axis plots $q_z$, the horizontal lines at $q_z=0$ and $q_z=2\pi$ correspond to  passages through the turning point at $\zmax$, while the horizontal line at $q_z=\pi$ indicates the turning point at $-\zmax$. Finally, at the horizontal lines at $q_z=\pi/2$ and $q_z=3\pi/2$ the orbit passes through the equator.

A striking feature of the plots in  Fig. \ref{fig:forceplotz1}-\ref{fig:forceplotz9} is that for the $t$, $r$, and $\phi$ components the GSF appears to be $\pi$ periodic in $q_z$ (rather than $2\pi$-periodic as one would naively expect). This feature is easily verified numerically. If we take decomposition in Fourier modes, this means only even multiples of $q_z$ appear in the exponents. A visually much less obvious feature is that  for the $z$ component of the GSF only odd multiples of $q_z$ appear in the exponents. Both of these features trace back to the original up/down symmetry of the Kerr background. In future iterations of the code, this feature can be used to speed up computation (and reduce memory usage) by sampling only half the range of $q_z$. This feature also explains the frequency oscillations seen in the time series of the $\phi$ component; in this component the polar variations dominate consequently, we see variations with a frequency $2\Omega_z$ in the time series.
 
\section{Conclusions and Outlook}
In this paper we have presented the first calculation of the first order gravitational self-force on generic Kerr geodesics. We have thus reached an important milestone in the  numerical GSF calculations needed to analyze EMRIs in LISA data.

However, there is still more work to be done. Although the calculations in this paper work fine as a proof of concept, the implementation is fairly slow. Major optimizations are needed in order to be be able to fill the $(a,p,e,\zmax)$ orbital parameter space as is needed to evolve inspirals \cite{Warburton:2011fk,Osburn:2015duj}. In particular, it seems we may have reached the limit of what is feasible in a \emph{Mathematica} implementation, which is great for prototyping new calculation methods such as this one, but not necessarily very efficient in the usage of CPU time and memory. A next step would be to implement in a more efficient compiled programming language.

The limited runs done for this paper involved only very modest eccentricities of $e=0.1$. LISA EMRIs are expected to have eccentricities of up to $e\lesssim0.8$. Such calculations would require significantly more modes and therefore computation resources. Our code for equatorial orbits has reached such eccentricities \cite{wigglespaper}, but at high computational cost. Without a more efficient implementation it seems infeasible to reach such high eccentricities with the current generic orbit code.

An exciting phenomenon that can be studied using the GSF on inclined eccentric orbits is the occurrence of orbital resonances \cite{Flanagan:2010cd}. These resonances are linked to an inspiral making a sudden jump in the constants of motion \cite{Flanagan:2010cd}. In principle, all information about these jumps can be extracted from the GSF at the moment of resonance \cite{vandeMeent:2013sza}. In particular, we should be able to settle the question whether there are contributions to the jumps from the conservative GSF that cannot be obtained from the fluxes \cite{Isoyama:2013yor}. There is also an intimate link with the question of integrability of the conservative GSF \cite{Vines:2015efa}. Calculating the $\psi_4$ generated by a resonant orbit, would require some minor modifications of our code as shown in \cite{Flanagan:2012kg}. However, beyond that all our methods should work almost identically.

The ability to calculate the GSF on inclined orbits further opens the door for the calculation for a slew of new (quasi-)invariants. These include the shift of the innermost stable spherical orbit \cite{Fujita:2016igj}, the equatorial limit of the nodal precession, the periapsis shift of spherical orbits, and of course the Detweiler redshift. The calculations of these quantities will require knowledge of the gauge completion \cite{gaugecompletion}, and will be pursued in future work.

\section*{Acknowledgements}
The author would like to acknowledge Scott Hughes for providing valuable verification data for testing the used Teukolsky solver for generic orbits. He also wishes to thank Leor Barack for a number of useful discussions. The author was supported by European Union's Horizon 2020 research and innovation programme under grant agreement No 705229. The numerical results in this paper were obtained using the IRIDIS High Performance Computing Facility at the University of Southampton.
 
\raggedright
\bibliography{../bib/journalshortnames,../bib/meent,../bib/commongsf,gengsf}

\end{document}

%% file: balancelaw.tab
\begingroup
%\squeezetable
\begin{table*}[t]
\caption{Numerical test of the balance law for a selection of strong field orbits. In each entry the first row gives $\mr^{-1}(\langle\d{\mathcal{C}_\mathrm{flux}^\pI}{\pt}\rangle + \langle\d{\mathcal{C}_\mathrm{flux}^\pH}{\pt}\rangle)$ (with $\mathcal{C}$ either $\nE$, $\nL$, or $\nQ$) calculated from the asymptotic values of $\psi_4$. The second row gives $-\mr^{-1}\langle\d{\mathcal{C}^\mathrm{GSF}}{t}\rangle$. These independently calculated quantities agree up to the estimated error level, providing a strong consistency check of the radiation self-force formalism, our numerical implementation, and error estimates. The brackets $(.)$ at the end of values indicated the estimated uncertainty on the last digit(s) (e.g. $ 1.234(5)\times 10^{-6}$ indicates $ 1.234\times 10^{-6} \pm 5\times 10^{-9}$).
}\label{tab:blaw}
\[\begin{array}{d{2}d{2}d{2}d{2}|l|l|l}
\toprule
 a & p & e & z& \mr^{-1}\avg{\d{\nE}{\pt}} & \mr^{-1}\avg{\d{\nL}{\pt}} & \mr^{-1}\avg{\d{\nQ}{\pt}}  \Ts\Bs\\
\hline
 0.9 & 10 & 0.1 & 0.1 &
\begin{array}[t]{l}
 -6.06082909932(12)\times 10^{-5} \\
 -6.060823(6)      \times 10^{-5} \\
\end{array}
&
\begin{array}[t]{l}
 -1.922697948770(32)\times 10^{-3} \\
 -1.922694(2)\times 10^{-3} \\
\end{array}
 & 
\begin{array}[t]{l}
 -1.25834727896(52)\times 10^{-4} \\
 -1.258348(1) \times 10^{-4} \\
\end{array}
 \Ts
\\
\hline
 0.9 & 10 & 0.1 & 0.3 &
\begin{array}[t]{l}
 -6.0940549476(48) \times 10^{-5} \\
 -6.094073(6)      \times 10^{-5} \\
\end{array}
&
\begin{array}[t]{l}
 -1.86084947343(96)\times 10^{-3} \\
 -1.860854(2)      \times 10^{-3} \\
\end{array}
 & 
\begin{array}[t]{l}
 -1.1433881349(42)\times 10^{-3} \\
 -1.143378(1) \times 10^{-3} \\
\end{array}
 \Ts\\
 \hline
 0.9 & 10 & 0.1 & 0.5 &
\begin{array}[t]{l}
 -6.16961426(18)\times 10^{-5} \\
 -6.16962(6)      \times 10^{-5} \\
\end{array}
&
\begin{array}[t]{l}
 -1.726403385(41)\times 10^{-3} \\
 -1.72640(2)\times 10^{-3} \\
\end{array}
 & 
\begin{array}[t]{l}
 -3.24433301(17)\times 10^{-3} \\
 -3.24429(3) \times 10^{-3} \\
\end{array}
 \Ts\\
\hline
 0.9 & 10 & 0.1 & 0.7 &
\begin{array}[t]{l}
 -6.3149587(13)\times 10^{-5} \\
 -6.31488(6)   \times 10^{-5} \\
\end{array}
&
\begin{array}[t]{l}
 -1.48663121(23)\times 10^{-3} \\
 -1.48661(2)	\times 10^{-3} \\
\end{array}
 & 
\begin{array}[t]{l}
 -6.6139045(19)\times 10^{-3} \\
 -6.6137(3)	 \times 10^{-3} \\
\end{array}
 \Ts\\
\hline
 0.9 & 10 & 0.1 & 0.9 &
\begin{array}[t]{l}
 -6.63479732(11) \times 10^{-5} \\
 -6.6341(7)      \times 10^{-5} \\
\end{array}
&
\begin{array}[t]{l}
 -1.015120160(27)\times 10^{-3} \\
 -1.01517(4)     \times 10^{-3} \\
\end{array}
 & 
\begin{array}[t]{l}
 -1.185295472(21)\times 10^{-2} \\
 -1.18535(8)	 \times 10^{-2} \\
\end{array}
 \Ts\\
\botrule
\end{array}\]
\end{table*}
\endgroup